\documentclass[aps,prc,onecolumn,groupedaddress,showpacs,showkeys]{revtex4}

\usepackage{graphicx}
\usepackage{epsfig,latexsym,amssymb}
\usepackage{amsmath}
\usepackage{bm}
\usepackage{slashed}
\usepackage{subfigure}

\newcommand{\be}{\begin{equation}}
\newcommand{\ee}{\end{equation}}
\newcommand{\bea}{\begin{eqnarray}}
\newcommand{\eea}{\end{eqnarray}}
\newcommand{\bem}{\begin{multline}}
\newcommand{\eem}{\end{multline}}
\newcommand{\beg}{\begin{gather}}
\newcommand{\eeg}{\end{gather}}

\newcommand{\as}{\alpha_s}

\newcommand{\ben}{\begin{eqnarray*}}
\newcommand{\een}{\end{eqnarray*}}

\newcommand{\msbar}{\mu_{\overline {\text{MS}}}}

\graphicspath{{Figs/}}

\begin{document}

\begin{flushright}
LU TP 16-XX\\
July 2016
\vskip1cm
\end{flushright}

\title{Testing the running coupling $k_{T}$-factorization formula for the inclusive gluon production} 
\author{F.O. Dur\~aes$^{1}$, A.V. Giannini$^{2}$, V.P. Gon\c{c}alves$^{3,4}$ and  F.S. Navarra$^2$}
\affiliation{$^1$ Curso de F\'{\i}sica, Escola de Engenharia, Universidade Presbiteriana Mackenzie\\
CEP 01302-907, S\~{a}o Paulo, Brazil \\
$^2$ Instituto de F\'{\i}sica, Universidade de S\~{a}o Paulo,
C.P. 66318,  05315-970 S\~{a}o Paulo, SP, Brazil\\
$^3$ Department of Astronomy and Theoretical Physics, Lund University, SE-223 62 Lund, Sweden \\  
$^{4}$ High and Medium Energy Group, Instituto de F\'{\i}sica e Matem\'atica,  Universidade Federal de Pelotas\\
Caixa Postal 354,  96010-900, Pelotas, RS, Brazil.\\
}

\begin{abstract}
The inclusive gluon production at midrapidities is described in the Color Glass Condensate formalism using the $k_T$ - factorization formula, which was derived at fixed coupling constant considering the scattering of a dilute system of partons with a dense one. Recent analysis  demonstrated that this approach provides a satisfactory description of the experimental data for the  inclusive hadron production in $pp/pA/AA$ collisions. However, these studies are based on the fixed coupling $k_T$ - factorization formula, which does not take into account the running coupling corrections, which are important to set the scales present in the cross section. In this paper we consider the running coupling corrected  $k_T$ - factorization formula  conjectured some years ago and investigate the impact of the running coupling corrections on the observables. In particular, the pseudorapidity distributions and charged hadrons multiplicity are calculated considering $pp$, $dAu/pPb$ and $AuAu/PbPb$ collisions at RHIC and LHC energies. We compare the corrected running coupling predictions with those obtained using the original $k_T$ - factorization assuming a fixed coupling or a prescription for the inclusion of the running of the coupling. Considering  the Kharzeev - Levin - Nardi  unintegrated gluon distribution and a simplified model for the nuclear geometry, we demonstrate that the distinct predictions are similar for the pseudorapidity distributions in $pp/pA/AA$ collisions and for the charged hadrons multiplicity in $pp/pA$ collisions. On the other hand, the running coupling corrected  $k_T$ - factorization formula predicts a smoother energy dependence for $dN/d\eta$ in $AA$ collisions.

\end{abstract}
\maketitle
\vspace{1cm}

\section{Introduction}

\date{\today}

The understanding of  inclusive hadron production in hadronic collisions is an important challenge for the theory of the strong interactions, 
since these processes  are expected to be dominated by small transverse momentum exchange. In general, nonperturbative approaches and/or 
phenomenological models based on soft physics (e.g. Reggeon approach) are used to study  hadron production with a satisfactory description of 
the experimental data. However, a shortcoming of these  approaches is that they are not based on  quarks and gluons and have no clear connection to  
the Quantum Chromodynamics (QCD). The QCD dynamics at high energies and large nuclei predicts the 
formation of a new state of matter, called Color Glass Condensate (For a review see Refs. \cite{review1}), characterized by the saturation scale 
$Q_s$, which is the typical momentum scale in the hadron wave function. The presence of this scale, which increases with energy and the atomic number,  
allows to treat  hadron production on  a solid theoretical basis, where perturbative methods can be applied. In the last years, the framework of the CGC 
approach have been used to describe with success the experimental data for the  hadron production. In high energy collisions we  
expect  to observe the transition from a linear description of the QCD dynamics, based on the DGLAP~\cite{dglap} and/or BFKL~\cite{bfkl} evolution equations, 
to a nonlinear description based on the Color Glass Condensate formalism \cite{hdqcd}. The transition between these two regimes is determined by the 
saturation scale $Q_s$, which grows with the energy and atomic number.  In the last years a comprehensive phenomenological analysis has been carried out 
to understand the HERA, RHIC and LHC data \cite{review1,hdqcd}.  Several theoretical studies have improved the CGC formalism by the inclusion of higher order 
corrections \cite{balnlo,kovnlo,iancunlo,lappinlo}. In particular, the running coupling corrections to the kernel of the Balitsky - Kovchegov (BK) equation 
\cite{bk} were calculated in Refs. \cite{balnlo,kovnlo}, with the solution being able to describe several observables at HERA, RHIC and LHC. More recently, the
 contributions of  large single and double transverse momentum logarithms have been resummed to all orders and included in the BK equation 
\cite{iancunlo,lappinlo}, with the resulting evolution equation being stable and generating   a physically meaningful evolution of the dipole amplitude. 
In addition, the formalism of  single inclusive hadron production in the framework of the hybrid approach proposed in Ref. \cite{dhj} has been improved by the inclusion of  next-to-leading 
order (NLO) corrections in  Refs. \cite{stasto_nlo,armesto_nlo,watanabe},   and a  generalization to higher orders of the $k_T$ - factorization formalism for  
inclusive gluon production was conjectured in Ref. ~\cite{Kovchegov.Horowitz}. As demonstrated in Ref. \cite{watanabe}, the NLO corrections of the hybrid 
formalism bring a better agreement of the predictions with the LHC and RHIC data on forward hadron production. In contrast, the impact of
the higher order corrections in the  $k_T$ - factorization formalism on  observables   is still unknown.  This is the subject of the present paper.

The $k_T$ - factorization formalism of gluon production in the central rapidity region (where the wave functions of both colliding particles are probed 
in the small-x regime) has been proposed in Ref. \cite{glr} and has been derived in the  leading $\log (1/x)$ and fixed coupling approximations in Ref. \cite{KT}, 
considering the scattering of a dilute parton system on a dense one (See also Ref. \cite{braun}). In a series of papers \cite{KLN}, Kharzeev, Levin and Nardi 
(KLN) have studied particle production at midrapidities in $pp/pA/AA$ collisions in terms of the $k_T$ - factorization formalism.  
They have assumed that the main properties of hadron production, as for example the energy, rapidity and transverse momentum dependence, are determined in 
the initial stage of the collision by the interaction between gluons with transverse momentum of the order of the saturation scale $Q_s$. The presence of 
this scale  regularizes the infrared behavior of the parton transverse momentum distributions and justifies a perturbative approach to the process. Since the  
basic predictions of the KLN approach have been qualitatively confirmed by  RHIC and LHC data, several authors have improved the KLN formalism in order to 
obtain a quantitative description of these data. In particular, in Refs.  \cite{amir,albadum,Albacete,dumitru,tribedy}, different models of the unintegrated 
gluon distribution and/or distinct treatments of the nuclear geometry have been considered. Although the $k_T$ - factorization formula has been derived 
assuming that $\alpha_s$ is a constant,  these different phenomenological studies have considered the running of the coupling constant and verified that it 
leads to an improvement of the agreement between theory and data. However, analysing in more detail these distinct predictions, we can observe that they were 
obtained using different choices of scale for the running coupling constant. Such uncertainty is expected in a leading order calculation, where the scales of 
the couplings are not known. Consequently, the inclusion of  running coupling corrections in  inclusive gluon production is an important step  to obtain 
quantitative predictions with higher accuracy. In Ref. ~\cite{Kovchegov.Horowitz}, the authors have calculated the running coupling corrections for the 
lowest - order gluon production cross section using the scale - setting prescription due to Brodsky, Lepage and Mackenzie (BLM) \cite{blm}. They found that 
the resulting cross section is expressed in terms of seven factors with running couplings, instead of the three  present in the fixed coupling calculation. 
In particular, two of these running couplings run with complex - valued momentum scales, which are complex conjugates of each other, implying real production 
cross sections. Finally, this calculation  fixes the scales of the running coupling constants appearing in the cross section. Based on these results for 
lowest - order gluon production, the authors have proposed a running coupling corrected $k_T$ - factorization formula, which is expected to be valid in the 
same regime as the original fixed order formula. Although the proof of this formula is still  an open question, it is expected that the resulting formula 
may still be a good approximation for the exact answer. Such expectation motivates the phenomenological analysis  performed in this paper. In what follows we 
will compare the predictions of the running coupling $k_T$ - factorization formula with those obtained assuming a fixed coupling constant and two different 
prescriptions for the inclusion of the running coupling in the leading order formula.  In all calculations we will use the same  model of the unintegrated gluon 
distribution  and we will use the same prescription for hadronization. Moreover, we will consider a simplified treatment of the nuclear geometry. This procedure allows us to  estimate more precisely the impact of the running coupling 
corrections on the $k_T$ - factorization formula. Such aspects surely can and should be improved in a quantitative comparison of the formalism with the 
experimental data. However, we believe that our main conclusions will  not be strongly modified.

This paper is organized as follows.  In the next Section we will present a brief review of the $k_T$ - factorization formalism and discuss the different 
prescriptions for the treatment of the coupling constant which will be considered in our analysis.  In Section \ref{res},  the KLN model of the unintegrated 
gluon distribution will be presented as well as the basic formulas for the calculation of the observables. Moreover, we will compute the pseudo - rapidity 
and energy distributions measured in hadron production in $pp/pA/AA$ collisions at RHIC and LHC energies. We will use  the running coupling 
$k_T$ - factorization formula and compare its predictions  with those obtained assuming a fixed coupling constant and two different prescriptions for the 
inclusion of the running coupling in the leading order formula. Finally, in Section \ref{conc} we summarize our main conclusions.

\section{Inclusive gluon production in the $k_{T}$-factorization formalism}
\label{formulas}

In this Section we will discuss  inclusive gluon production in the $k_{T}$-factorization formalism.  
Before presenting  the main formulas, a comment is in order. As described in the Introduction, the cross section of this process was proposed in Ref. \cite{glr} 
and proven in Ref. \cite{KT} (See also Ref. \cite{braun})  considering the scattering of a dilute partonic system on a dense one at fixed coupling constant 
and in leading $\log (1/x)$ approximation. Consequently, its application is well justified for  gluon production at midrapidity in $pA$ collisions. On the 
other hand, in the case of $pp$ and $AA$ collisions at high energies, the gluon jet at midrapidities is produced by the interaction of two dense systems. 
In such cases, factorization breaking effects are expected to become significant \cite{raju}, modifying the basic $k_{T}$-factorization formulas. However, 
the magnitude of these corrections  is still subject of intense debate and its contribution in the kinematical range probed by the LHC is not well known. The 
fact that the $k_{T}$-factorization formalism allows us to obtain a very good description of the current data, suggests that these corrections are not 
large and that this formalism can be considered a reasonable approximation for the treatment of  gluon production in $pp$ and $AA$ collisions at central 
rapidities. Therefore, in what follows, we will apply the $k_{T}$-factorization formalism to  $pp$, $pA$ and $AA$ collisions.

Let us consider the production of a gluon with momentum $k_T$ at rapidity $y$ in a collision between the hadrons $h_1$ and $h_2$, with $h_i = p$ or $A$. 
In the $k_{T}$-factorization formalism, the differential cross section for this process will be given by \cite{KT}
\begin{align}\label{ktfact}
  \frac{d^3 \sigma}{d^2 k_T \, dy} \, = \, \frac{2}{C_F} \,
  \frac{1}{{\bm k}^2} \, \int d^2 q \, \as \, \phi_{h_1} ({\bm q}, y) \, \phi_{h_2}
  ({{\bm k} - \bm q}, Y-y)\,,
\end{align}
where $Y$ is the total rapidity interval of the collision, $C_{F} = (N_{c}^{2}-1)/2N_{c}$ 
and boldface variables denote transverse plane vectors, ${\bm k} = k_T = (k^{1},k^{2})$. Moreover,  $\phi_{h_i} (x_i,\bm q)$ denotes the so-called unintegrated 
gluon  distribution, which represents the probability to find a gluon with momentum fraction $x_i$ and transverse momentum $\bm q$ in the hadron $h_i$. This 
 distribution can be expressed as follows
\begin{align}\label{ktglueA}
  \phi_{h_i} ({\bm k}, y) \, = \, \frac{C_F}{\as \, (2 \pi)^3} \, \int d^2 b \,
d^2 r \, e^{- i {\bm k} \cdot {\bm r}} \ \nabla^2_r \, \mathcal{N}^G_{h_i} ({\bm r},
{\bm b}, y) \,\,,
\end{align}
with $\mathcal{N}^G_{h_i} ({\bm r}, {\bm b}, y)$ being the dipole - hadron $h_i$ forward scattering amplitude for a gluon dipole of transverse size 
$\bm r$ and  $\bm b$  the impact parameter of the scattering. The behavior of $\mathcal{N}^G$ at large rapidities (small - $x$) is directly related 
to the QCD dynamics at high energies. In the general case, it will be given by the solution of the JIMWLK evolution equation \cite{cgc}, but in the 
large-$N_c$ limit it can be expressed in terms of the solution of the BK equation \cite{bk} for the quark dipole - hadron forward scattering amplitude. 
As the numerical solution of these equations including the impact parameter dependence is still very challenging \cite{levin,stasto_biernat,berger}, in the studies 
of  gluon production using $k_T$ - factorization formula the authors have introduced simplifying assumptions about the impact parameter dependence of the 
phenomenological models for the unintegrated gluon distributions (or about the quark dipole scattering amplitude), which are based on  CGC physics and have 
their  parameters constrained by  experimental data \cite{amir,albadum,Albacete,dumitru,tribedy}. 

In the $k_T$ - factorization formula, Eq. (\ref{ktfact}), the cross section is proportional to the coupling constant $\alpha_s$, which was assumed to be 
constant in its derivation. Moreover, $\alpha_s$  appears also in the unintegrated gluon distribution, Eq. (\ref{ktglueA}). In the last years,  running 
coupling corrections have been calculated for the BK-JIMWLK evolution equations and this allows us to estimate the contribution of these corrections to 
$\mathcal{N}^G$. However, it is still not clear how to determine the momentum scale in $\alpha_s$ in Eq. (\ref{ktglueA}).   This has motivated  the generalization 
of  Eqs. (\ref{ktfact}) and  (\ref{ktglueA}) by the inclusion of  running coupling constant corrections \cite{amir,albadum,Albacete,dumitru,tribedy}. In general, 
these studies assume that the  factorized expression is preserved by these corrections and that the coupling constants in  
Eqs. (\ref{ktfact}) and  (\ref{ktglueA}) depend on different momentum scales. The choice of these scales is arbitrary and we found different choices in the 
literature. For example, in Ref. \cite{amir}, the authors assume that $\alpha_s = \alpha_s (k_T^2)$ in Eq. (\ref{ktfact}) and $\alpha_s = \alpha_s(Q^2_s(x_i))$ 
in Eq. (\ref{ktglueA}). On the other hand, in Ref. \cite{Albacete} it is assumed that  $\alpha_s = \alpha_s (Q^2)$ in Eq. (\ref{ktfact}), with 
$Q^{2} = \mbox{max}\{k^{2},(k - q)^{2}\}$,   and the scale of running coupling in Eq. (\ref{ktglueA}) is assumed to be equal to the transverse momentum of the 
gluon.  A common characteristic of these approaches is that they assume the leading order running coupling
\begin{align}
\label{alphas}
\as(Q^{2}) = \frac{12 \pi}{\beta_{0}\ln(Q^{2}/\Lambda_{QCD}^{2})}\,; \quad\quad\quad \beta_{0} = 33 - 2n_{f}
\end{align}
where $\Lambda_{QCD}$ is a non-perturbative scale and $n_{f}$ is the number 
of fermions. Moreover, very often a smooth freezing of the coupling at  low scales is assumed. For example, in Ref. \cite{amir}  the strong coupling is taken   
to be  $\as (Q^{2}) = 0.5$ when $Q^{2} \le 0.8$ GeV$^{2}$.
As pointed out in Refs. \cite{amir,Albacete}, the inclusion of  running coupling corrections improves the description of the experimental data. However, 
as discussed in detail in Ref. \cite{Kovchegov.Horowitz}, it is not clear that  Eq. (\ref{ktfact}) will keep its factorized form after the inclusion of 
these corrections. In order to clarify this aspect, the authors of \cite{Kovchegov.Horowitz} have estimated the running coupling corrections in  lowest-order gluon 
production cross section, finding that three factors of fixed coupling in lowest-order expression should be replaced by seven running couplings, with the 
new structure being called the {\it septumvirate} of couplings. Two scales of the couplings are complex-valued, but given the structure of the expression, 
the cross section is real. Moreover, the cross section is symmetric in the  parton momentum scales.  In Ref. \cite{Kovchegov.Horowitz} the authors have 
proposed a generalization of  the lowest-order expression to higher-orders, which  includes the small-$x$ evolution. They proposed the following expression for 
the running coupling corrected $k_T$ - factorization formula
\begin{eqnarray}\label{rc_fact}
  \frac{d^3 \sigma}{d^2 k_T \, dy} \, = \, \frac{2 \, C_F}{\pi^2} \,
  \frac{1}{{\bm k}^2} \, \int d^2 q \ {\overline \phi}_{h_1} ({\bm q}, y)
  \, {\overline \phi}_{h_2} ({\bm k} - {\bm q}, Y-y) \, \frac{\as \left(
      \Lambda_\text{coll}^2 \, e^{-5/3} \right)}{\as \left( Q^2 \,
      e^{-5/3} \right) \, \as \left( Q^{* \, 2}\, e^{-5/3} \right)} \,\,,
\end{eqnarray}
with the unintegrated gluon distribution functions defined by
\begin{equation}\label{rc_ktglueA}
  {\overline \phi}_{h_i} ({\bm k}, y) \, = \as \phi_{h_i} ({\bm k}, y) \, = \, \frac{C_F}{(2 \pi)^3} \,
  \int d^2 b \, d^2 r \, e^{- i {\bm k} \cdot {\bm r}} \ \nabla^2_r \,
  \mathcal{N}_{h_i} ({\bm r}, {\bm b}, y)\,.
\end{equation}
where $\Lambda_\text{coll}^2$ is a collinear infrared cutoff and the momentum scale $Q$ is given by
\begin{align}\label{Qscale}
  \ln \frac{Q^2}{\msbar^2} \, & = \, \frac{1}{2} \, \ln \frac{{\bm
      q}^2 \, ({\bm k} - {\bm q})^2}{\msbar^4} - \frac{1}{4 \, {\bm
      q}^2 \, ({\bm k} - {\bm q})^2 \, \left[ ({\bm k} - {\bm q})^2 -
      {\bm q}^2 \right]^6} \, \Bigg\{ {\bm k}^2 \, \left[ ({\bm k} -
    {\bm q})^2 - {\bm q}^2 \right]^3 \notag \\[5pt]
  & \quad \; \times \, \bigg\{ \left[ \left[({\bm k}-{\bm
        q})^2\right]^2 - \left({\bm q}^2\right)^2 \right] \, \left[
    \left({\bm k}^2\right)^2 + \left(({\bm k}-{\bm q})^2 - {\bm
        q}^2\right)^2 \right] + 2 \, {\bm k}^2 \, \left[ \left({\bm
        q}^2\right)^3 - \left[({\bm k}-{\bm q})^2\right]^3
  \right] \notag \\[5pt]
  & \quad \; - {\bm q}^2 \, ({\bm k} - {\bm q})^2 \left[ 2 \,
    \left({\bm k}^2\right)^2 + 3 \, \left[({\bm k} - {\bm q})^2 - {\bm
        q}^2\right]^2 - 3 \, {\bm k}^2 \, \left[({\bm k} - {\bm q})^2
      + {\bm q}^2\right] \right] \, \ln
  \left( \frac{({\bm k} - {\bm q})^2}{{\bm q}^2} \right) \bigg\} \notag \\[5pt]
  & + \, i \, \left[ ({\bm k} - {\bm q})^2 - {\bm q}^2 \right]^3 \, \,
  \bigg\{ {\bm k}^2 \, \left[ ({\bm k} - {\bm q})^2 - {\bm q}^2\right]
  \, \left[ {\bm k}^2 \, \left[ ({\bm k} - {\bm q})^2 + {\bm
        q}^2\right] - \left({\bm q}^2\right)^2 -
    \left[({\bm k}-{\bm q})^2\right]^2 \right] \notag \\[5pt]
  & \quad \; + {\bm q}^2 \, ({\bm k} - {\bm q})^2 \, \left( {\bm k}^2
    \, \left[ ({\bm k} - {\bm q})^2 + {\bm q}^2\right] - 2 \,
    \left({\bm k}^2\right)^2 - 2 \, \left[ ({\bm k} - {\bm q})^2 -
      {\bm q}^2\right]^2 \right) \, \ln \left( \frac{({\bm k} - {\bm
        q})^2}{{\bm q}^2} \right) \bigg\} \notag \\[5pt]
  & \quad \; \times \, \sqrt{2 \, {\bm q}^2 \, ({\bm k} - {\bm q})^2 +
    2 \, {\bm k}^2 \, ({\bm k} - {\bm q})^2 + 2 \, {\bm q}^2 \, {\bm
      k}^2 - \left({\bm k}^2\right)^2 - \left({\bm q}^2\right)^2 -
    \left[({\bm k}-{\bm q})^2\right]^2} \Bigg\}\,\,,
\end{align}
with $\mu_{\bar{MS}}^2$ being the renormalization scale in the $\bar{MS}$ scheme. Differently from Eq. (\ref{ktfact}), in the corrected expression 
all scales in the coupling constants are  specified. Moreover, it has the expected behaviours for ${\bm q} \rightarrow 0$ and ${\bm q} \rightarrow {\bm k}$ \cite{Kovchegov.Horowitz}.  In Ref. \cite{Kovchegov.Horowitz} the authors claim that Eq. (\ref{rc_fact}), like 
Eq. (\ref{ktfact}), is valid  both in the linear and non-linear regimes of the QCD dynamics. However, as emphasized there, only exact 
calculations can check the validity of this conjecture.  We expect Eq. (\ref{rc_fact})  to be a good approximation for the 
exact answer. This expectation motivates a phenomenological study using the running  coupling corrected $k_T$ - factorization formula.

\section{Results}
\label{res}

In this section we will compare the predictions of the running coupling corrected $k_T$ - factorization formula,  given by Eq. (\ref{rc_fact}) and 
denoted CF hereafter, with those derived using the original formula, Eq. (\ref{ktfact}). In the latter, we will calculate the inclusive gluon 
production cross section assuming a fixed value for the coupling constant (denoted FC), and also assuming that the couplings run according to the 
prescriptions proposed in Refs. \cite{Albacete} and \cite{amir},  denoted hereafter by RC1 and RC2, respectively. In order to clarify the impact of 
the running coupling corrections in the $k_T$ - factorization formula we will make the following approximations: (a) we will disregard 
the impact parameter dependence of the unintegrated gluon distributions and consider only minimum bias collisions assuming that  $A_{eff} = 20\, (18.5)$ 
for Pb (Au); (b) we will assume the validity of the principle of Local Parton - Hadron Duality (LPDH), which implies that the form of the rapidity 
distribution for the hadron spectrum  differs from the gluon spectrum only by a numerical factor. This introduces an effective mass $m_h$ 
(it will be always equal to $0.350$ GeV) which  incorporates nonperturbative effects  and (c) we will use a single model of the unintegrated 
gluon distribution, namely  the KLN model \cite{KLN}, which encodes the basic aspects of the nonlinear QCD dynamics, and is given by 
\begin{eqnarray}\label{phiKLN}
  \phi_{KLN} ({\bm k}, y) \, &=& \frac{2 C_F}{3\,\pi^2\,\as} \,, \quad k \le Q_{s}  \\
                            &=& \frac{2 C_F}{3\,\pi^2\,\as} \frac{Q_{s}^{2}}{k^{2}} \,, \quad k > Q_{s}\,,
\end{eqnarray}
where the saturation scale is given by $Q_{s}^{2} = A_{eff}^{1/3} Q_{0}^{2}(x_{0}/x)^{\lambda}$ with 
$Q_{0} = 1$ GeV, $x_{0} = 3 \times 10^{-4}$ and $\lambda = 0.288$ \cite{GBW.model}. As in previous works \cite{KLN,amir}, we will multiply 
$\phi_{KLN}$ by a factor $(1 - x)^4$ as prescribed by quark couting rules \cite{Brodsky:1973kr,Matveev:1973ra} in order to simulate the behavior of 
the distribution at large $x$ ($x \rightarrow 1$). 
All these approximations can and should be improved in a quantitative calculation of the observables. However, we believe that our simplified analysis 
can help us to get an insight on how  the running coupling corrections (included in the $k_T$ - factorization formula) change the observables.
In what follows we will calculate the inclusive  hadron production cross section, which is given by
\begin{equation}
\frac{d^3N}{d\eta d^2k_T} = \frac{K}{\sigma_s} \, h(y,k_{T},m_{h}) \cdot \frac{d^3 \sigma}{d^2 k_T \, dy} \,\,,
\label{yield}
\end{equation} 
where $\eta$ is the pseudorapidity and $h(y,k_{T},m_{h})$ is the Jacobian  
for the conversion from 
rapidity to pseudorapidity, which is given by 
\begin{align}\label{etatoy}
h(y,k_{T},m_{h}) = \sqrt{1 - \frac{m_{h}^{2}}{m_{T}^{2}\cosh^{2}y }}\,,
\end{align}
with $m_T^2 = k_T^2 + m_h^2$. 
The  $K$-factor incorporates in an effective way the contribution of higher order corrections, of possible effects not included in the CGC formalism 
and also the uncertainty in  the conversion of partons to hadrons. Moreover, $\sigma_s$ is the average interaction area. 
As in previous works \cite{amir}, we will correct the kinematics due to the presence of the mass scale $m_h$, replacing 
$k_{T}\rightarrow \sqrt{k_{T}^{2} + m_{h}^{2}}$ in the definition of the Bjorken-$x$ variable and also in the factor $\frac{1}{{\bm k}^2}$ appearing 
in Eqs. (\ref{ktfact}) and (\ref{rc_fact}). Moreover,  we will choose $\alpha_s = 0.25$ in the fixed coupling calculations and 
$(N_c,\,n_f,\,\Lambda_{QCD}) = (3, 3, 0.240 \, \mbox{GeV})$ in the RC1 and RC2 predictions. In the case of the corrected expression we will assume   
$\as \left( \Lambda_\text{coll}^2 \, e^{-5/3} \right) = 0.25$ and the value of  $\msbar$ will be fixed by requiring that  $\msbar^{2}\, e^{5/3} = 
\Lambda_{QCD}^{2}$. Finally, the normalization factor ${K}/{\sigma_s}$ will be treated as a free parameter to be fixed by the comparison with the 
experimental data at a given energy and/or rapidity.

\begin{figure}[t]
\begin{center}
\subfigure[ ]{
\includegraphics[width=0.475\textwidth]{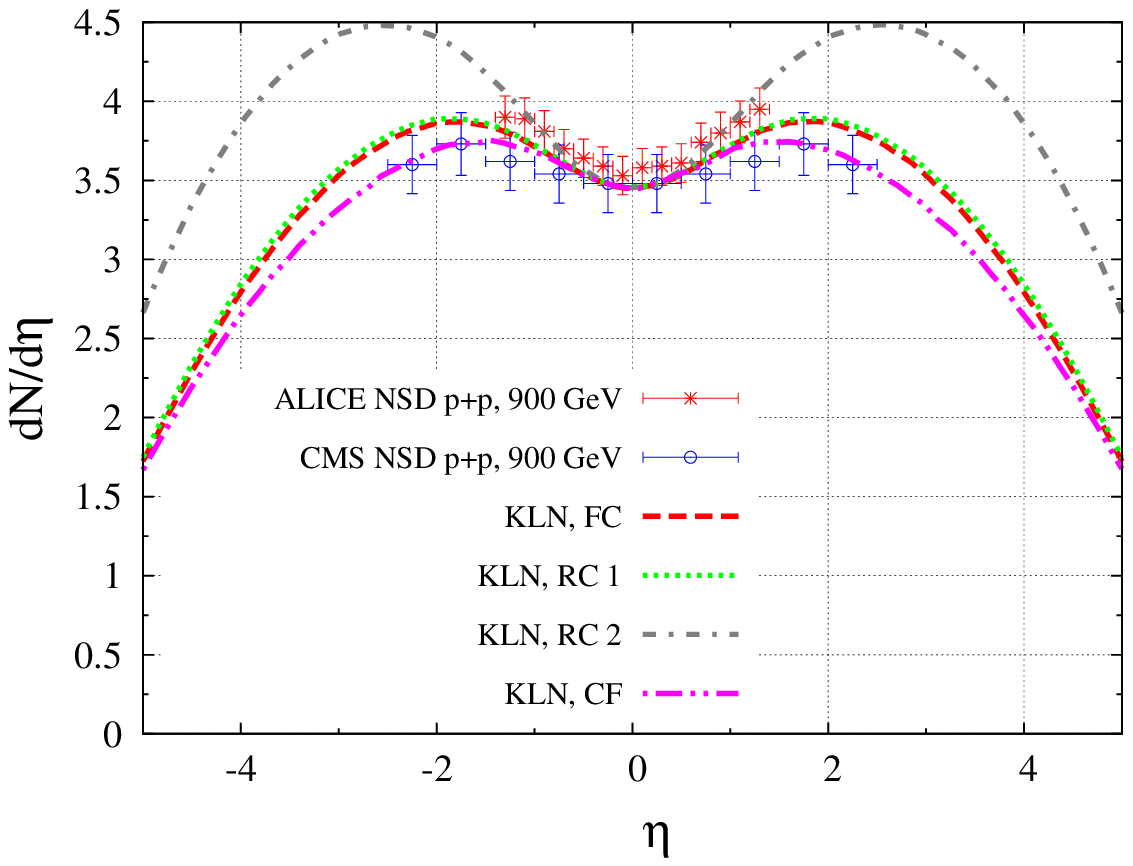}}
\subfigure[ ]{
\includegraphics[width=0.475\textwidth]{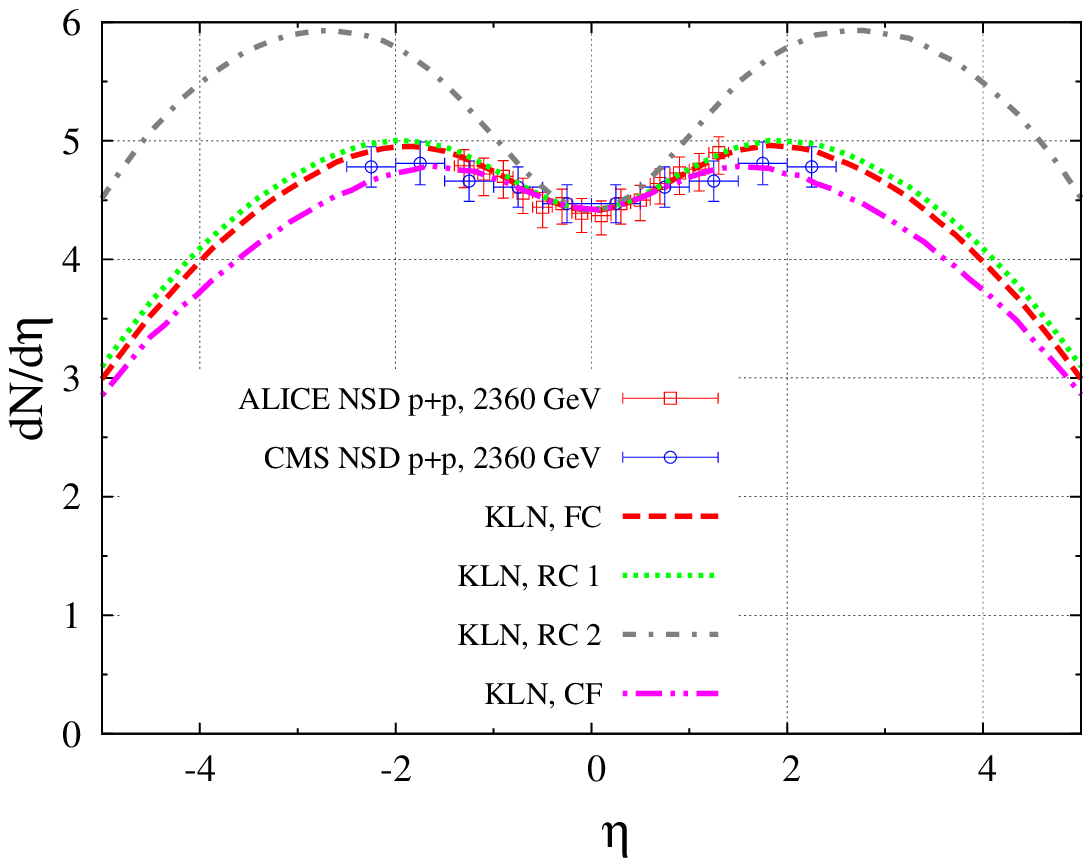}}
\subfigure[ ]{
\includegraphics[width=0.475\textwidth]{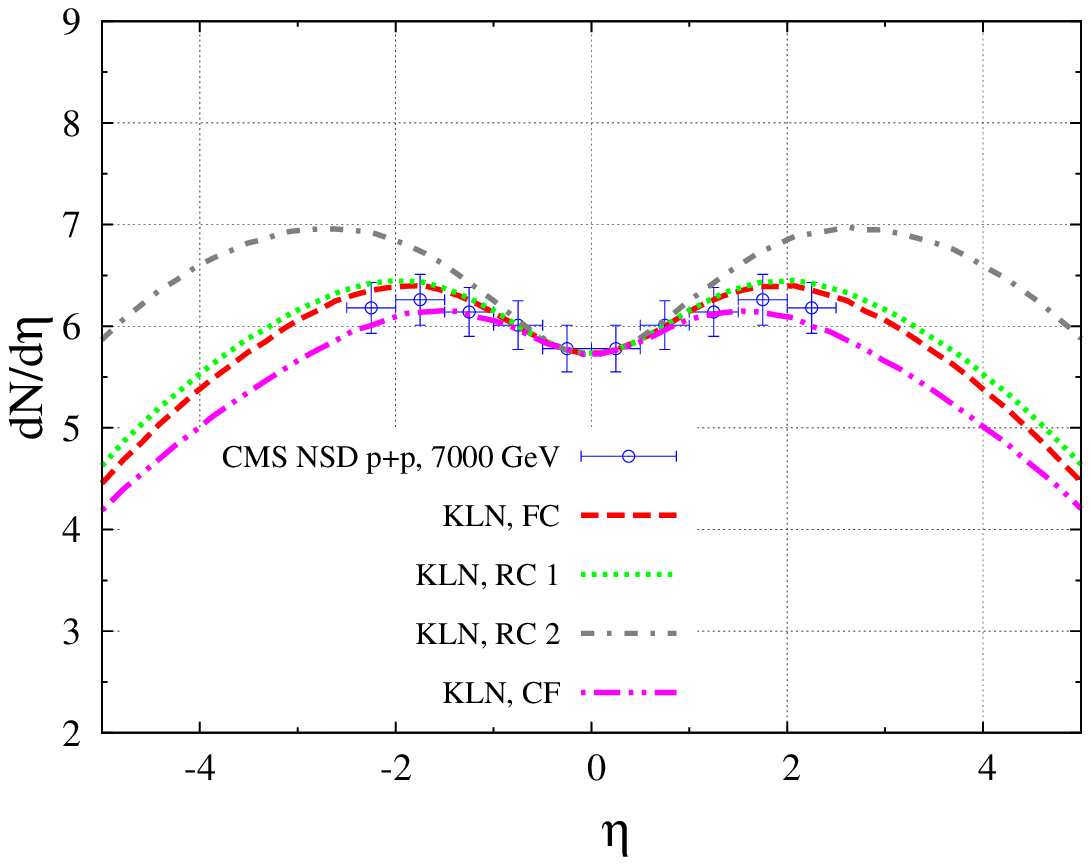}}
\end{center}
\vskip -0.80cm
\caption{Pseudorapidity distributions measured in  inclusive hadron production in $pp$ collisions for different values of $\sqrt{s}$:  
(a) $0.9$ TeV, (b) $2.36$ TeV and  (c) $7$ TeV. Data are from Refs. \cite{Aamodt:2010ft,Khachatryan:2010xs,Khachatryan:2010us}. }
\label{Fig:dNdetam045datafreenormpp}
\end{figure}

\begin{figure}[t]
\begin{center}
\subfigure[ ]{
\includegraphics[width=0.475\textwidth]{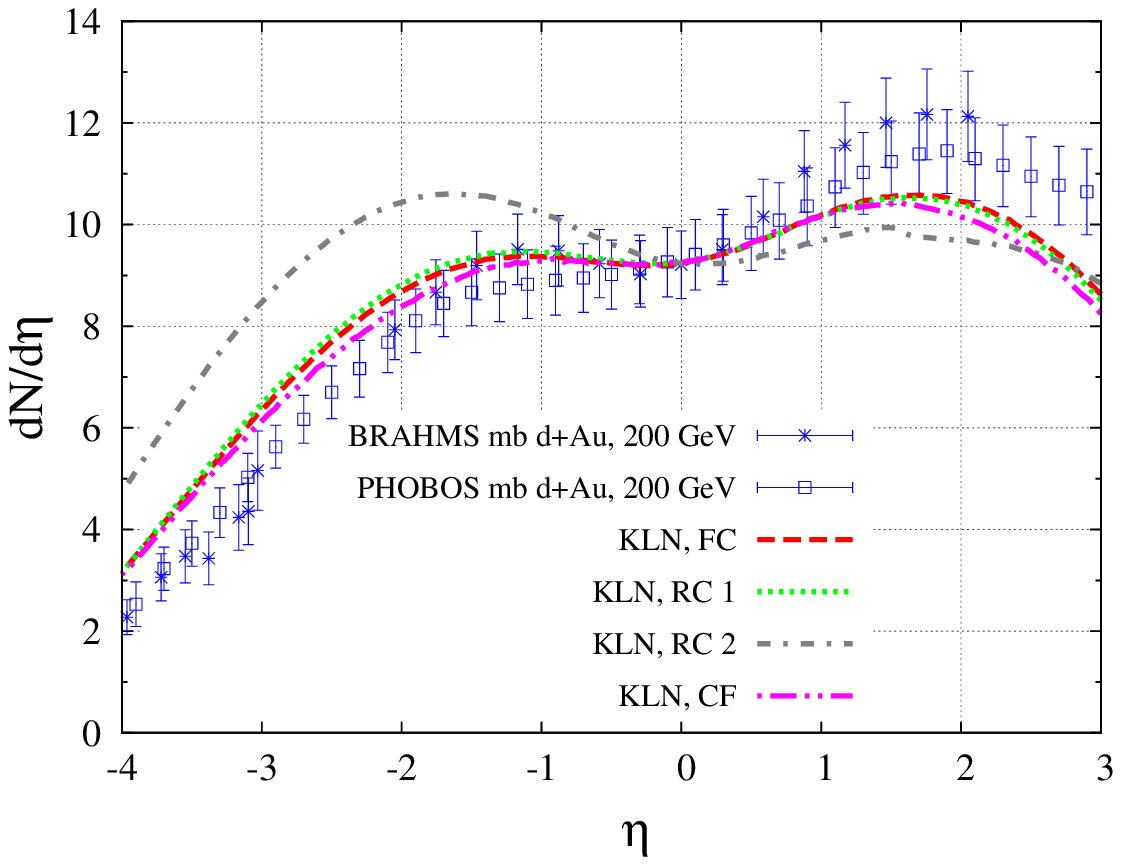}}
\subfigure[ ]{
\includegraphics[width=0.475\textwidth]{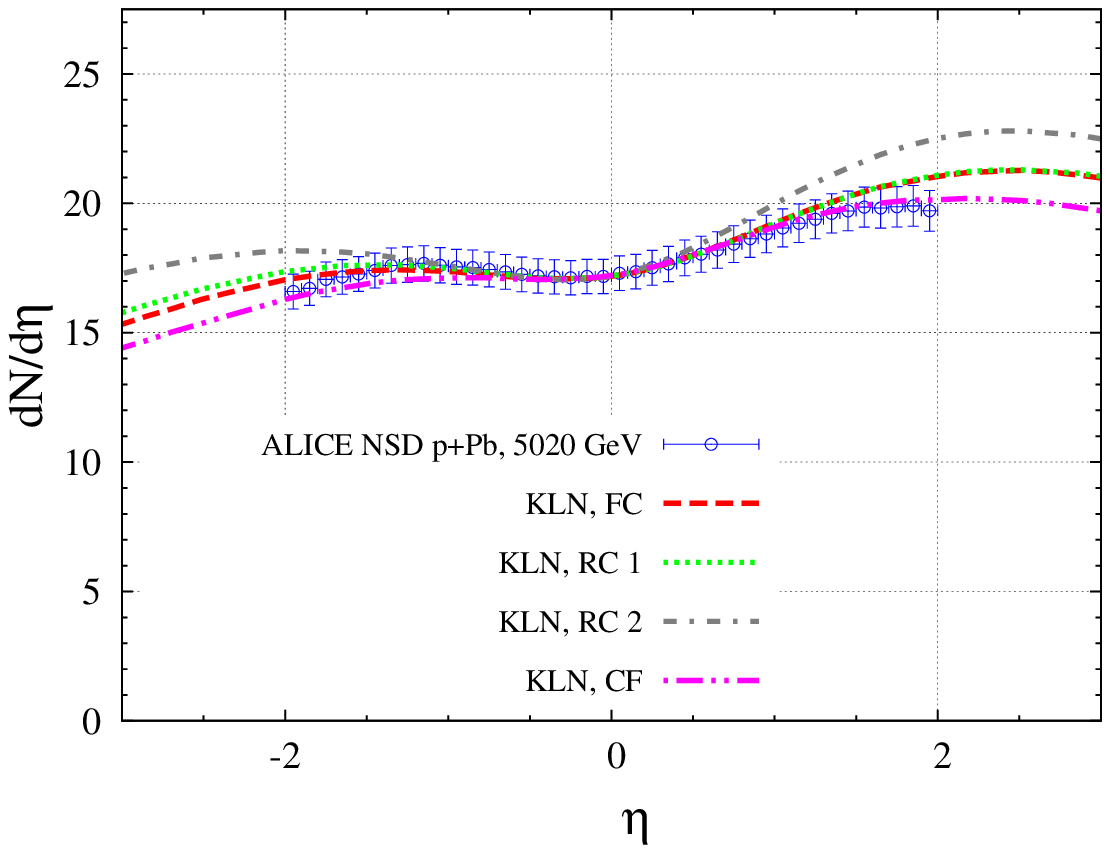}}
\end{center}
\vskip -0.80cm
\caption{Pseudorapidity distributions measured in  inclusive hadron production in (a) $dAu$ collisions ($\sqrt{s} = 0.2$ TeV) and (b) 
$pPb$ collisions ($\sqrt{s} = 5.02$ TeV). Data are from Refs. \cite{Arsene:2004cn,Alver:2010ck,ALICE:2012xs}.}
\label{Fig:dNdetamh045datapA}
\end{figure}

In Fig. \ref{Fig:dNdetam045datafreenormpp} we present our results for the pseudorapidity distributions obtained in  $pp$ collisions at different 
center-of-mass energies. The normalization of the different curves, given by the factor $K/\sigma_s$ in Eq. (\ref{yield}), has been fixed at each 
energy in order to reproduce the experimental data at $\eta = 0$.  The FC and RC1 predictions are very similar, differing of the other results at 
large $\eta$, whereas the RC2 curve exhibits the steepest rise (and fall) with the pseudorapidity.  The CF formula yields  a reasonable description of the 
$pp$ data. We have verified that the necessary  change of the  normalization between   $\sqrt{s} = 0.9$ TeV and  
7 TeV is smaller  than 1.0 \% in the case of the CF predictions. On the other hand, for the other predictions, the change was larger than 19\%.  

In Fig. \ref{Fig:dNdetamh045datapA} we present our results for inclusive hadron production in  $dAu$ collisions ($\sqrt{s} = 0.2$ TeV) and in $pPb$ 
collisions ($\sqrt{s} = 5.02$ TeV). The normalization of the $dAu$ results was chosen so as to describe  the data on the deuteron side and in the 
$\eta = 0$ region (within the error bars),  simultaneously. On the other hand, in the $pPb$ case, we normalize our predictions in such way that they 
reproduce the data at $\eta = 0$. As observed in our analysis of the $pp$ results, the different models predict a distinct  behavior  with $\eta$, 
with the FC and RC1  curves being similar and the RC2 one predicting a steeper dependence.  The CF  prediction is able to describe the experimental 
data in a large range of pseudorapidities. The discrepancy appearing at large $\eta$ in $dAu$ collisions can be attributed to the simplified treatment  
of  nuclear geometry used in our calculations. Concerning the change of the normalization $K/\sigma_s$ necessary to fit  data at different energies, 
we observe that the smallest change occurs for the CF predictions ($\approx 17$ \%), while in the other predictions the change is of order of 40 \%. 
A possible interpretation of this result is that  the corrected formula captures important energy dependent higher - order contributions, since  the 
same model of  the unintegrated gluon distribution was used in all predictions.

\begin{figure}[t]
\begin{center}
\subfigure[ ]{
\includegraphics[width=0.475\textwidth]{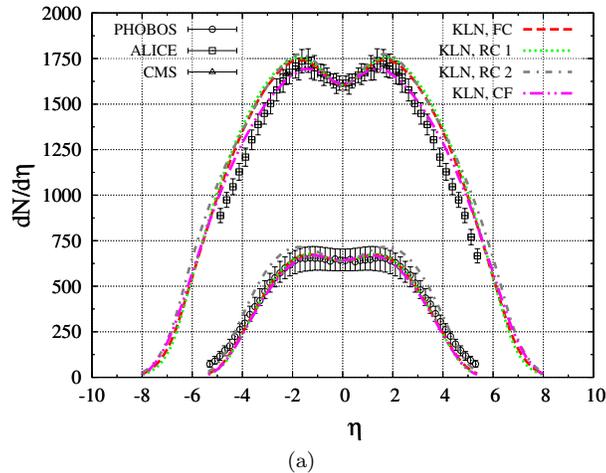}}
\end{center}
\vskip -0.80cm
\caption{Pseudorapidity distributions measured in  inclusive hadron production in $AuAu$ collisions at $\sqrt{s} = 0.2$ TeV (lower curves) and $PbPb$ 
collisions at $\sqrt{s} = 2.76$ TeV (upper curves). Data are from Refs. \cite{Back:2001bq,Back:2002wb,Abbas:2013bpa,Chatrchyan:2011pb}.}
\label{Fig:dNdetamh045dataAA}
\end{figure}

In Fig. \ref{Fig:dNdetamh045dataAA} we present our predictions for  $Au Au$ collisions at $\sqrt{s} = 0.2$ TeV and  $Pb Pb$ collisions at $\sqrt{s} = 2.76$ 
TeV. The normalization of the different curves has been fixed in order to reproduce the data at $\eta = 0$. The $\eta$ dependence of the  different curves  
is  similar, with the CF one providing a reasonable description of the experimental data. In contrast with the $pp$ and $pPb/dAu$ cases, in heavy ion 
collisions the required change of the normalization, when going from one energy to another,   is always large, even in the  CF case $[{\cal{O}}(25 \%)]$.

\begin{figure}[t]
\begin{center}
\subfigure[ ]{
\includegraphics[width=0.475\textwidth]{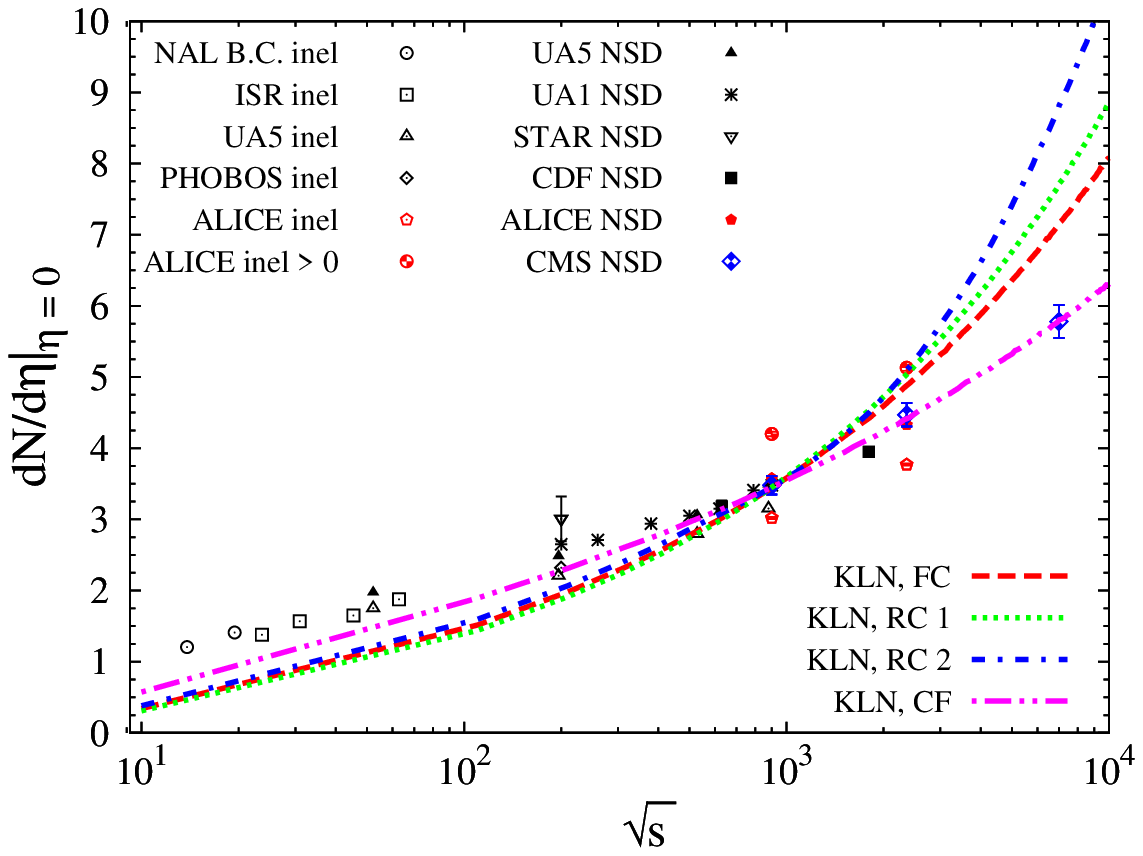}}
\subfigure[ ]{
\includegraphics[width=0.475\textwidth]{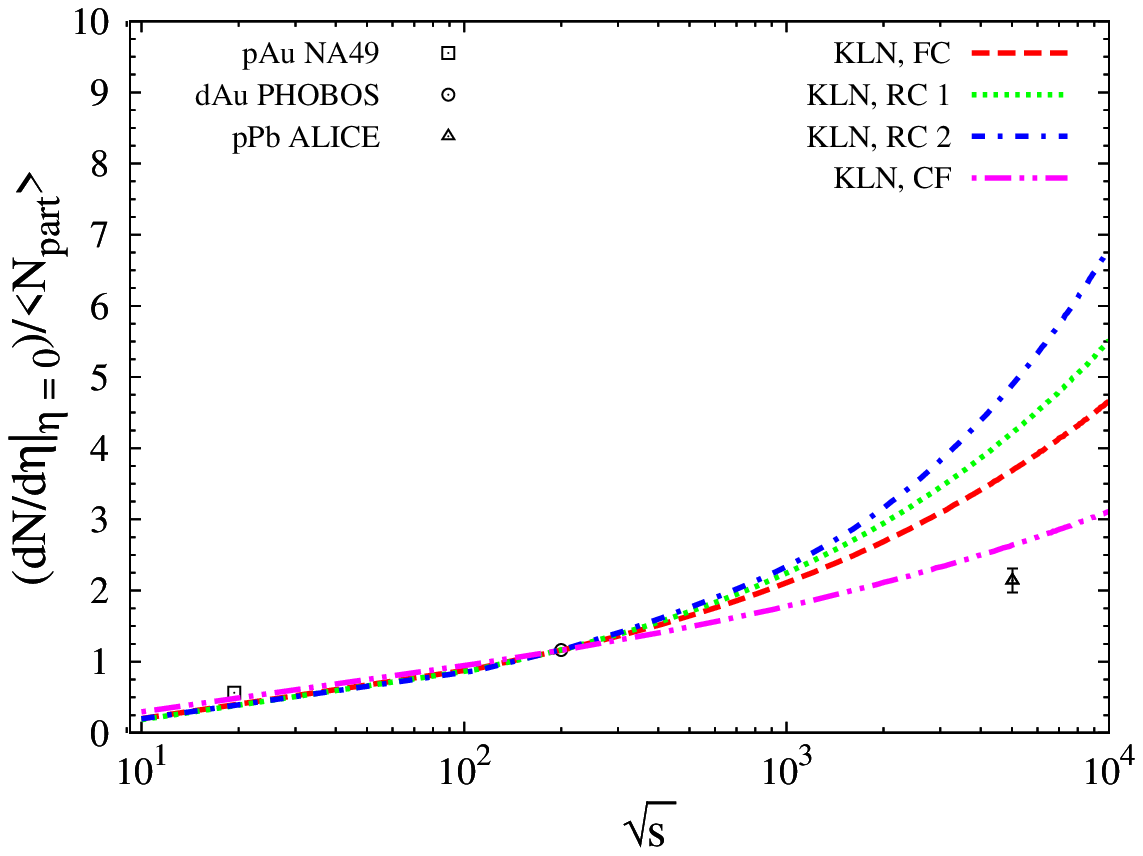}}
\subfigure[ ]{
\includegraphics[width=0.475\textwidth]{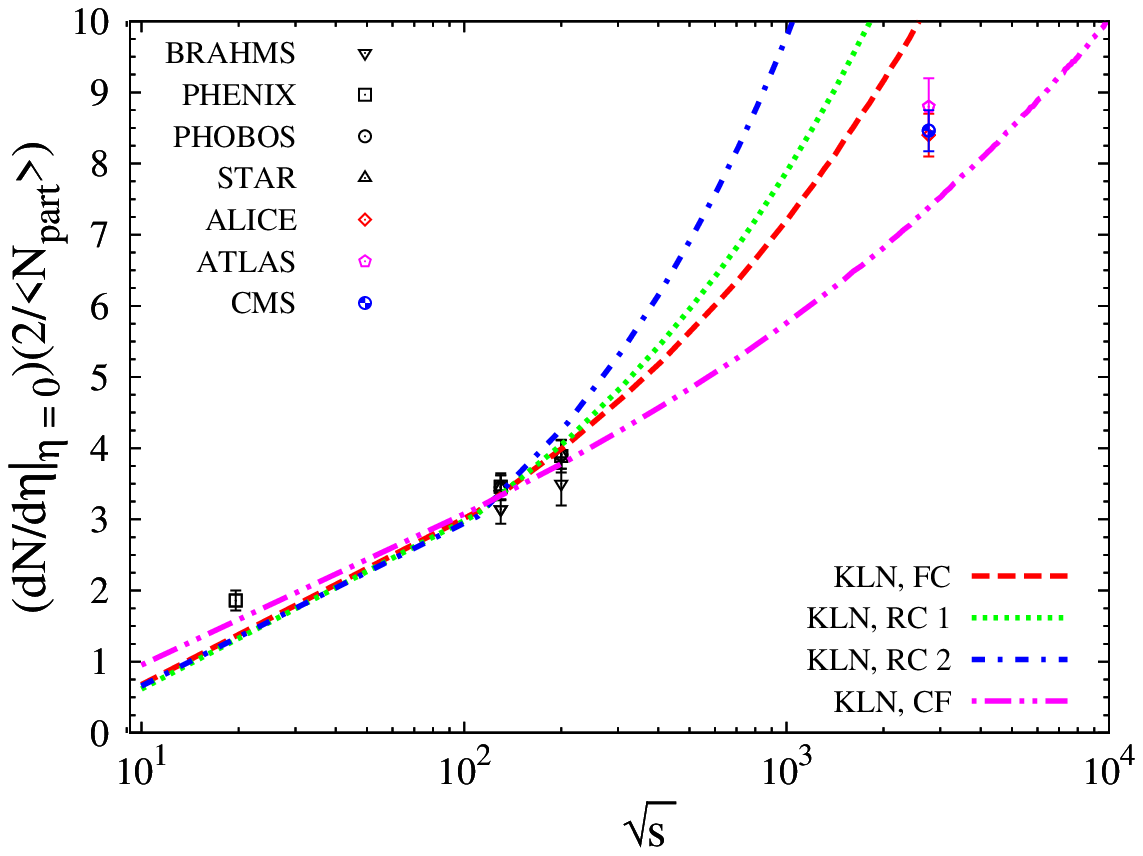}}
\end{center}
\vskip -0.80cm
\caption{Energy dependence of charged hadron multiplicities in the central region of rapidity $\eta = 0$.  (a) $pp$, (b) $pA$ and (c) $AA$ collisions. 
The $pA$ and $AA$ predictions  are  normalized by $N_{\rm part}$ and $2/\langle N_{\rm part}\rangle$, respectively. Data are from 
Refs.  \cite{Aamodt:2010ft,Khachatryan:2010xs,ALICE:2012xs,Chatrchyan:2011pb,Back:2002uc,
Whitmore:1973ri,Thome:1977ky,Alner:1986xu,Nouicer:2004ke,Aamodt:2010pp,
Albajar:1989an,Abelev:2008ab,Abe:1989td,Bearden:2001xw,Bearden:2001qq,
Adler:2004zn,Aamodt:2010cz,ATLAS:2011ag}.}
\label{Fig:dNdetavssqrtsmh045}
\end{figure}

Finally, let us compare the predictions of the different models for the energy dependence of charged hadron multiplicities at  $\eta = 0$. We will 
consider $pp$, $pA$ and $AA$ collisions, with the $pA$ and $AA$ predictions  being normalized by $\langle N_{\rm part}\rangle$ and 
$2/\langle N_{\rm part}\rangle$, respectively, where $\langle N_{\rm part}\rangle$ is the average number of participants. We use the values of $\langle N_{\rm part}\rangle$ 
given in Ref. \cite{ALICE:2012xs,Back:2002uc} for minimum bias $pPb$ collisions and the $3\%$ most central $AA$ collisions. 
As we are interested in the energy dependence of the predictions, we will fix the  normalization factors using the experimental data on $dN/d\eta$ 
in $pp$ collisions at $\sqrt{s} = 0.9$ TeV,  $dAu$ collisions at $\sqrt{s} = 0.2$ TeV and  $AuAu$ collisions at $\sqrt{s} = 0.13$ TeV. The predictions for 
higher energies will be parameter free.  As can be seen in  Fig. \ref{Fig:dNdetavssqrtsmh045} the CF curve presents a slower growth with the energy in 
comparison with the predictions obtained using the 
original $k_T$ - factorization formula. One have that using a simplified model for the unintegrated gluon distribution and a crude treatment of the nuclear geometry, the corrected $k_T$ - factorization formula implies a satisfactory description of the $pp$ and $pA$ data. In particular,  the CF predictions describe quite well the experimental data from $pp$ collisions at high and low 
energies, in contrast to the other approaches that using the same inputs fail to reproduce data at $\sqrt{s} < 0.9$ TeV.   Moreover, the corrected formula provides a prediction for the  $pA$ case 
that is closer to the experimental data. In contrast, the CF predictions underestimate the  $AA$ data for high energies, which could indicate that for such systems a more precise treatment of the unintegrated gluon distribution and nuclear geometry is fundamental to describe the data. Surely, such aspects should be investigated in the future.

\section{Conclusions}
\label{conc}

In the phenomenology of the CGC, the $k_T$ - factorization formula, is one of the most important tools. It  was originally derived assuming a 
fixed coupling constant and a collision between a dilute and a dense system. The effect of  running coupling corrections on the 
$k_T$ - factorization formula was addressed in Ref. \cite{Kovchegov.Horowitz}, where a corrected expression was proposed. The study of the 
implications of these corrections on the observables was analysed, for the first time, in this paper. Considering simplistic assumptions for the 
nuclear geometry and for the unintegrated gluon distribution, we have estimated the cross section for  inclusive hadron production in $pp/pA/AA$ 
collisions and we  have compared our results with the predictions derived from the original formula, from a fixed coupling and also from two 
different prescriptions for the scale choice in the running coupling constant. We demonstrated that  the impact of these corrections on the observables is small, with the predictions of the distinct approaches for the pseudorapidity 
distributions and charged hadron multiplicities being similar. In particular, we verified that the predictions of the corrected formula yield  
a satisfactory description of the experimental data. The main difference arises in the energy dependence of the observables,  with the corrected formula predicting  a weaker energy dependence.
Our results motivate more robust calculations considering a realistic unintegrated gluon distribution and a precise treatment of the nuclear geometry.  Surely these aspects deserve to be investigated in more detail in the future. However, we believe that the exploratory study performed in this paper shed light on the basic implications of the corrected 
running coupling $k_T$ - factorization formula and that the basic conclusions which emerge from this analysis will remain valid.

\begin{acknowledgments}
The authors would like to thank A. Dumitru  by useful discussions during the initial stage of this study and by useful comments in a first version of the manuscript. This work was  partially financed by the 
Brazilian funding agencies CNPq, CAPES, FAPERGS and FAPESP.

\end{acknowledgments}


\begin{thebibliography}{99}

\bibitem{review1}
  N.~Armesto {\it et al.},
  J.\ Phys.\ G {\bf 35} (2008) 054001;  
  C.~A.~Salgado {\it et al.},
  J.\ Phys.\ G {\bf 39}, 015010 (2012); J.~L.~Albacete {\it et al.},
  Int.\ J.\ Mod.\ Phys.\ E {\bf 22}, 1330007 (2013)

\bibitem{dglap} 
Yu. Dokshitzer, Sov. Phys. JETP {\bf 46}, 1649 (1977); V.N. Gribov and L.
N. Lipatov, Sov. Nucl. Phys. {\bf 15}, 438 (1972);
G. Altarelli and G. Parisi, Nucl. Phys. {\bf B126}, 298 (1977).

\bibitem{bfkl} 
L. N. Lipatov, Sov. J. Nucl. Phys. {\bf 23}, 338 (1976); E. A. Kuraev,
L. N. Lipatov and V. S. Fadin, Sov. Phys. JETP {\bf 45}, 199 (1977);
I. I. Balitsky and L. N. Lipatov, Sov. J. Nucl. Phys. {\bf 28}, 822 (1978).


\bibitem{hdqcd} F.~Gelis, E.~Iancu, J.~Jalilian-Marian and R.~Venugopalan, Ann.\ Rev.\ Nucl.\ Part.\ Sci.\  {\bf 60}, 463 (2010);E.~Iancu and R.~Venugopalan,  
              arXiv:hep-ph/0303204;  H.~Weigert,  Prog.\ Part.\ Nucl.\ Phys.\  {\bf 55}, 461 
              (2005); J.~Jalilian-Marian and Y.~V.~Kovchegov,
              Prog.\ Part.\ Nucl.\ Phys.\  {\bf 56}, 104 (2006); J.~L.~Albacete and C.~Marquet, 
              Prog.\ Part.\ Nucl.\ Phys.\  {\bf 76}, 1 (2014).

\bibitem{balnlo}
  I.~Balitsky,
  Phys.\ Rev.\  D {\bf 75}, 014001 (2007); 
  I.~Balitsky and G.~A.~Chirilli,
  Phys.\ Rev.\  D {\bf 77}, 014019 (2008).



\bibitem{kovnlo}
  Y.~V.~Kovchegov and H.~Weigert,
  Nucl.\ Phys.\  A {\bf 784}, 188 (2007); 
  Nucl.\ Phys.\  A {\bf 789}, 260 (2007);   Y.~V.~Kovchegov, J.~Kuokkanen, K.~Rummukainen and H.~Weigert,
  Nucl.\ Phys.\  A {\bf 823}, 47 (2009).


\bibitem{iancunlo} 
  E.~Iancu, J.~D.~Madrigal, A.~H.~Mueller, G.~Soyez and D.~N.~Triantafyllopoulos,
  Phys.\ Lett.\ B {\bf 750}, 643 (2015)

\bibitem{lappinlo} 
  T.~Lappi and H.~Mantysaari,
  Phys.\ Rev.\ D {\bf 93}, 094004 (2016)

\bibitem{bk} 
I. Balitsky, {Nucl. Phys.} {\bf B463}, 99 (1996); Y. V. Kovchegov,
{Phys. Rev. D} {\bf 60}, 034008 (1999); {\it ibid}. {\bf 61}, 074018 (2000).

\bibitem{dhj}
A.~Dumitru, A.~Hayashigaki and J.~Jalilian-Marian,
  Nucl.\ Phys.\ A {\bf 765}, 464 (2006)


\bibitem{stasto_nlo} 
  G.~A.~Chirilli, B.~W.~Xiao and F.~Yuan,
  Phys.\ Rev.\ Lett.\  {\bf 108}, 122301 (2012); A.~M.~Stasto, B.~W.~Xiao and D.~Zaslavsky,
  Phys.\ Rev.\ Lett.\  {\bf 112}, 012302 (2014)
  
  \bibitem{armesto_nlo} 
  T.~Altinoluk, N.~Armesto, G.~Beuf, A.~Kovner and M.~Lublinsky,
  Phys.\ Rev.\ D {\bf 91}, 094016 (2015)


\bibitem{watanabe} 
  K.~Watanabe, B.~W.~Xiao, F.~Yuan and D.~Zaslavsky,
  Phys.\ Rev.\ D {\bf 92}, 034026 (2015)



\bibitem{Kovchegov.Horowitz}
  W.~A.~Horowitz and Y.~V.~Kovchegov,
  Nucl.\ Phys.\ A {\bf 849}, 72 (2011).

\bibitem{glr} 
  L.~V.~Gribov, E.~M.~Levin and M.~G.~Ryskin,
  Phys.\ Rept.\  {\bf 100}, 1 (1983).

\bibitem{KT}
Y.~V.~Kovchegov and K.~Tuchin,
  Phys.\ Rev.\ D {\bf 65}, 074026 (2002)

\bibitem{braun} 
  M.~A.~Braun,
  Phys.\ Lett.\ B {\bf 483}, 105 (2000).

\bibitem{KLN} 
  D.~Kharzeev and M.~Nardi,
  Phys.\ Lett.\ B {\bf 507}, 121 (2001);  
  D.~Kharzeev and E.~Levin,
  Phys.\ Lett.\ B {\bf 523}, 79 (2001);
  D.~Kharzeev, E.~Levin and M.~Nardi,
  Phys.\ Rev.\ C {\bf 71}, 054903 (2005);
Nucl.\ Phys.\ A {\bf 747}, 609 (2005)

  
\bibitem{amir} 
  E.~Levin and A.~H.~Rezaeian,
  Phys.\ Rev.\ D {\bf 82}, 014022 (2010); 
  Phys.\ Rev.\ D {\bf 82}, 054003 (2010);  
  Phys.\ Rev.\ D {\bf 83}, 114001 (2011).

\bibitem{albadum} 
  J.~L.~Albacete and A.~Dumitru,
  arXiv:1011.5161 [hep-ph].

\bibitem{Albacete}
  J.~L.~Albacete, A.~Dumitru, H.~Fujii and Y.~Nara,
  Nucl.\ Phys.\ A {\bf 897}, 1 (2013).

\bibitem{dumitru} 
  A.~Dumitru, D.~E.~Kharzeev, E.~M.~Levin and Y.~Nara,
  Phys.\ Rev.\ C {\bf 85}, 044920 (2012).


\bibitem{tribedy} 
  P.~Tribedy and R.~Venugopalan,
  Nucl.\ Phys.\ A {\bf 850}, 136 (2011)
  Erratum: [Nucl.\ Phys.\ A {\bf 859}, 185 (2011)]; Phys.\ Lett.\ B {\bf 710}, 125 (2012)
  Erratum: [Phys.\ Lett.\ B {\bf 718}, 1154 (2013)]; B.~Schenke, P.~Tribedy and R.~Venugopalan,
  Phys.\ Rev.\ C {\bf 89}, 024901 (2014).

\bibitem{blm} 
  S.~J.~Brodsky, G.~P.~Lepage and P.~B.~Mackenzie,
  Phys.\ Rev.\ D {\bf 28}, 228 (1983).  
  
  
\bibitem{raju}
F.~Gelis, T.~Lappi and R.~Venugopalan,
  Phys.\ Rev.\ D {\bf 78}, 054019 (2008);
Phys.\ Rev.\ D {\bf 78}, 054020 (2008); Phys.\ Rev.\ D {\bf 79}, 094017 (2009).  
  
\bibitem{cgc}  J. Jalilian-Marian, A. Kovner, L. McLerran  and  H.
Weigert, Phys. Rev. D {\bf 55}, 5414 (1997); J. Jalilian-Marian, A. Kovner and  H.
Weigert, Phys. Rev. D {\bf 59}, 014014 (1999), {\it ibid.} {\bf 59}, 014015 (1999), {\it ibid.} {\bf 59}  034007 (1999);
A. Kovner, J. Guilherme Milhano and  H. Weigert,   Phys. Rev. D {\bf 62},  114005 (2000);
 H. Weigert, Nucl. Phys.  {\bf A703}, 823 (2002); E. Iancu, A. Leonidov and L. McLerran, Nucl.Phys.  {\bf A692} (2001) 583;  E. Ferreiro, E. Iancu, A. Leonidov and  L. McLerran, Nucl. Phys. {\bf A701}, 489 (2002).
  


\bibitem{levin}
  E.~Gotsman, M.~Kozlov, E.~Levin, U.~Maor and E.~Naftali,
  Nucl.\ Phys.\ A {\bf 742}, 55 (2004); A.~Kormilitzin and E.~Levin,
  Nucl.\ Phys.\ A {\bf 849}, 98 (2011); C.~Contreras, E.~Levin and I.~Potashnikova,
  Nucl.\ Phys.\ A {\bf 948}, 1 (2016).

\bibitem{stasto_biernat}  
  K.~J.~Golec-Biernat and A.~M.~Stasto,
  Nucl.\ Phys.\ B {\bf 668}, 345 (2003).

\bibitem{berger}
J.~Berger and A.~Stasto,
  Phys.\ Rev.\ D {\bf 83}, 034015 (2011); Phys.\ Rev.\ D {\bf 84}, 094022 (2011); JHEP {\bf 1301}, 001 (2013)
  

\bibitem{GBW.model} 
  K.~J.~Golec-Biernat and M.~Wusthoff,
  Phys.\ Rev.\ D {\bf 59}, 014017 (1998).
  

\bibitem{Brodsky:1973kr} 
  S.~J.~Brodsky and G.~R.~Farrar,
  Phys.\ Rev.\ Lett.\  {\bf 31}, 1153 (1973).


\bibitem{Matveev:1973ra} 
  V.~A.~Matveev, R.~M.~Muradian and A.~N.~Tavkhelidze,
  Lett.\ Nuovo Cim.\  {\bf 7}, 719 (1973).
  


\bibitem{Aamodt:2010ft} 
  K.~Aamodt {\it et al.} [ALICE Collaboration],
  Eur.\ Phys.\ J.\ C {\bf 68}, 89 (2010).

\bibitem{Khachatryan:2010xs} 
  V.~Khachatryan {\it et al.} [CMS Collaboration],
  JHEP {\bf 1002}, 041 (2010).

\bibitem{Khachatryan:2010us} 
  V.~Khachatryan {\it et al.} [CMS Collaboration],
  Phys.\ Rev.\ Lett.\  {\bf 105}, 022002 (2010).





\bibitem{Arsene:2004cn} 
  I.~Arsene {\it et al.} [BRAHMS Collaboration],
  Phys.\ Rev.\ Lett.\  {\bf 94}, 032301 (2005).

\bibitem{Alver:2010ck} 
  B.~Alver {\it et al.} [PHOBOS Collaboration],
  Phys.\ Rev.\ C {\bf 83}, 024913 (2011).

\bibitem{ALICE:2012xs} 
  B.~Abelev {\it et al.} [ALICE Collaboration],
  Phys.\ Rev.\ Lett.\  {\bf 110}, no. 3, 032301 (2013).



\bibitem{Back:2001bq} 
  B.~B.~Back {\it et al.} [PHOBOS Collaboration],
  Phys.\ Rev.\ Lett.\  {\bf 87}, 102303 (2001).


\bibitem{Back:2002wb} 
  B.~B.~Back {\it et al.},
  Phys.\ Rev.\ Lett.\  {\bf 91}, 052303 (2003).


\bibitem{Abbas:2013bpa} 
  E.~Abbas {\it et al.} [ALICE Collaboration],
  Phys.\ Lett.\ B {\bf 726}, 610 (2013).


\bibitem{Chatrchyan:2011pb} 
  S.~Chatrchyan {\it et al.} [CMS Collaboration],
  JHEP {\bf 1108}, 141 (2011).

\bibitem{Back:2002uc} 
  B.~B.~Back {\it et al.} [PHOBOS Collaboration],
  Phys.\ Rev.\ C {\bf 65}, 061901 (2002).






\bibitem{Whitmore:1973ri} 
  J.~Whitmore,
  Phys.\ Rept.\  {\bf 10}, 273 (1974).

\bibitem{Thome:1977ky} 
  W.~Thome {\it et al.} [Aachen-CERN-Heidelberg-Munich Collaboration],
  Nucl.\ Phys.\ B {\bf 129}, 365 (1977).

\bibitem{Alner:1986xu} 
  G.~J.~Alner {\it et al.} [UA5 Collaboration],
  Z.\ Phys.\ C {\bf 33}, 1 (1986).

\bibitem{Nouicer:2004ke} 
  R.~Nouicer {\it et al.} [PHOBOS Collaboration],
  J.\ Phys.\ G {\bf 30}, S1133 (2004).


\bibitem{Aamodt:2010pp} 
  K.~Aamodt {\it et al.} [ALICE Collaboration],
  Eur.\ Phys.\ J.\ C {\bf 68}, 345 (2010)
  doi:10.1140/epjc/s10052-010-1350-2
  [arXiv:1004.3514 [hep-ex]].


\bibitem{Albajar:1989an} 
  C.~Albajar {\it et al.} [UA1 Collaboration],
  Nucl.\ Phys.\ B {\bf 335}, 261 (1990).


\bibitem{Abelev:2008ab} 
  B.~I.~Abelev {\it et al.} [STAR Collaboration],
  Phys.\ Rev.\ C {\bf 79}, 034909 (2009).


\bibitem{Abe:1989td} 
  F.~Abe {\it et al.} [CDF Collaboration],
  Phys.\ Rev.\ D {\bf 41}, 2330 (1990).




\bibitem{Bearden:2001xw} 
  I.~G.~Bearden {\it et al.} [BRAHMS Collaboration],
  Phys.\ Lett.\ B {\bf 523}, 227 (2001).


\bibitem{Bearden:2001qq} 
  I.~G.~Bearden {\it et al.} [BRAHMS Collaboration],
  Phys.\ Rev.\ Lett.\  {\bf 88}, 202301 (2002).


\bibitem{Adler:2004zn} 
  S.~S.~Adler {\it et al.} [PHENIX Collaboration],
  Phys.\ Rev.\ C {\bf 71}, 034908 (2005)
  Erratum: [Phys.\ Rev.\ C {\bf 71}, 049901 (2005)].

\bibitem{Aamodt:2010cz} 
  K.~Aamodt {\it et al.} [ALICE Collaboration],
  Phys.\ Rev.\ Lett.\  {\bf 106}, 032301 (2011).

\bibitem{ATLAS:2011ag} 
  G.~Aad {\it et al.} [ATLAS Collaboration],
  Phys.\ Lett.\ B {\bf 710}, 363 (2012).

\end{thebibliography}
\end{document}